\documentclass{iopjournal}

\usepackage{amssymb}
\usepackage{graphicx}
\usepackage{color}
\usepackage{caption}
\usepackage{hyperref}
\usepackage{amsthm}
\usepackage{amsmath}
\usepackage{braket}
\usepackage{appendix}
\usepackage{subcaption}
\usepackage{comment}
\usepackage{bbold}
\usepackage{multirow}
\usepackage{booktabs}
\usepackage{physics}

\usepackage{quantikz}
\usepackage[utf8]{inputenc} 
\usepackage[T1]{fontenc}
\DeclareUnicodeCharacter{2009}{\,}
\begin{document}

\articletype{Article type} 

\title{Topological frustration and quantum resources}

\author{Alberto Giuseppe Catalano$^1$$^2$$^*$\orcid{0000-0002-6643-0738}, Gianpaolo Torre$^3$\orcid{0000-0002-4274-6463}, Salvatore Marco Giampaolo$^2$\orcid{0000-0003-4647-8758} and Fabio Franchini$^2$\orcid{0000-0002-3429-8189}}

\affil{$^1$Dipartimento di Fisica e Astronomia "G. Galilei" \& Padua Quantum Technologies Research Center, Universit{\`a} degli Studi di Padova, Italy I-35131, Padova, Italy}

\affil{$^2$INFN, Sezione di Padova, via Marzolo 8, I-35131, Padova, Italy}

\affil{$^3$Institut Ruder Boškovi\'c, Bijeni\v{c}ka cesta 54, Zagreb 10000, Croatia}

\affil{$^*$Author to whom any correspondence should be addressed.}

\email{albertogiuseppe.catalano@unipd.it}


\begin{abstract}
Although in general boundary conditions do not affect the bulk properties of a system, some of them are special and defy such expectation. This is the case, for instance, of those inducing geometrical frustration in a classical magnet. Recently, the study of such settings in quantum systems (dubbed topological frustration) has uncovered peculiar features, interesting both from a fundamental and technological point of view. In this work, we present and discuss the behavior of several quantum resources in presence of TF, namely the (disconnected) entanglement entropy and the non-stabilizerness Renyi entropy. We will show that, compared to their non-frustrated counterparts, TF adds a distinct contribution to these resources, due to a stable, delocalized, topological excitation. Remarkably, this contribution can be calculated analytically, due to its similarities with that of a W-state.
\end{abstract}

\section{Introduction}

Quantum mechanics (QM) has played a central role in shaping modern science and technology over the past century. 
This role has become even more prominent in recent decades, as QM provides the fundamental framework for a rapidly expanding class of emerging technologies. 
Indeed, the current period is often described as the \emph{second quantum revolution}~\cite{PRXQuantum.1.020101, Preskill2018quantumcomputingin}, reflecting the transition from exploiting individual quantum effects to the controlled manipulation of complex quantum systems. 
Fields such as quantum information, quantum communication, and quantum computing have reached a remarkable level of maturity, driving the development of novel quantum-based technologies. 
While some of these technologies have already achieved commercial relevance, others, most notably quantum computing, are advancing at an unprecedented pace~\cite{RevModPhys.95.045005, Cerezo2021, Cao2019, Abbas2024}.

A crucial ingredient underlying this progress is the rapidly increasing ability to engineer, manipulate, and probe quantum many-body systems at the microscopic level~\cite{Acin2018, Riedel2017}. 
This capability has enabled the realization of \emph{synthetic quantum matter}, introducing a new paradigm in which complex quantum phenomena can be emulated in highly controllable and tunable experimental platforms~\cite{Browaeys2020, RevModPhys.91.015005, ArguelloLuengo2024, RevModPhys.91.015006, Henriet2020quantumcomputing, Zippilli2014}. 
Synthetic quantum matter encompasses a wide range of experimental realizations, including ultracold atoms trapped in optical lattices~\cite{RevModPhys.80.885}, trapped-ion systems~\cite{RevModPhys.93.025001}, and superconducting qubits implemented in circuit QED architectures~\cite{RevModPhys.93.025005}. 
These systems are deliberately engineered to reproduce specific many-body Hamiltonians and dynamical processes, rather than relying on naturally occurring materials with comparable properties, thereby enabling systematic investigations of regimes that are often inaccessible in conventional solid-state settings~\cite{Browaeys2020, RevModPhys.86.153}. 
As such, they provide a natural bridge between fundamental many-body physics and the development of quantum technologies.

\begin{figure}[t!]
	\centering
	\includegraphics[width=.6\columnwidth]{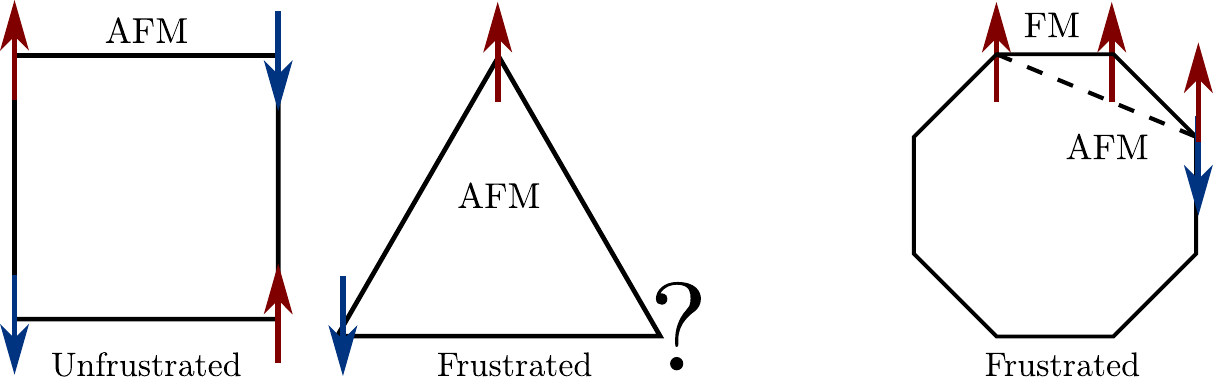}
	\caption{(Left) Illustration of geometric frustration through simple geometry. 
    The square lattice admits a N\'eel-ordered configuration in which all antiferromagnetic bonds are simultaneously satisfied and is therefore unfrustrated. 
    In contrast, the triangular plaquette does not allow all AFM interactions to be minimized at the same time: assigning alternating spins on two vertices inevitably forces the third one to violate one AFM bond. 
    (Right) Example of frustration induced by competing interactions. 
    Here, a next-to-nearest AFM coupling favours arrangements incompatible with the nearest-neighbor interaction, regardless of whether it is FM or AFM. 
    Since the two interactions cannot be simultaneously satisfied, the system becomes frustrated.}
	\label{fig:Topology1}
\end{figure}

Within this context, frustrated quantum systems are attracting growing attention for the peculiarity of their properties~\cite{Balents2010, Lacroix2011}. 
The concept of frustration was originally introduced in classical statistical mechanics to describe systems in which it is impossible to simultaneously minimize all constraints imposed by the Hamiltonian~\cite{Toulouse1986, Sadoc1999, Diep2012}. 
In magnetic systems, for example, frustration arises when spin configurations cannot be arranged so as to satisfy all interaction terms at once. 
Two paradigmatic mechanisms leading to frustration are geometric constraints, as encountered in non-bipartite lattices, and competing interactions promoting incompatible orders.
These two mechanism are depicted in Fig.~\ref{fig:Topology1}.
Canonical examples include antiferromagnetic models on triangular lattices~\cite{Wannier1950, Stephenson1970}, as well as one-dimensional systems with competing couplings such as the $J_1$--$J_2$ model~\cite{PhysRevB.54.9862} or the ANNNI chain~\cite{Beccaria2007, Selke1988, Torre_2025}.

In quantum many-body systems, however, the notion of frustration must be treated with particular care. 
Due to the intrinsically non-commuting nature of local operators, and with the exception of special cases such as frustration-free Hamiltonians~\cite{Movassagh2016, Tong2021} or systems at factorization points~\cite{Giampaolo2008, Giampaolo2009, Giampaolo2010}, most quantum models exhibit some degree of frustration~\cite{Wolf2003, Giampaolo2011, Marzolino2013, Giampaolo2015}. 
In accordance with common usage in the literature, throughout this work the term \emph{frustration} is restricted to effects originating from lattice geometry, competing interactions, or boundary conditions, thereby preserving its classical operational meaning~\cite{Toulouse1986, Vannimenus1977}.

Among the various sources of frustration, a particularly subtle and intrinsically global form is provided by \emph{topological frustration} (TF). 
TF arises from topological constraints that cannot be resolved locally, but instead are enforced by the global structure of the system. 
A prototypical realization is found in antiferromagnetic spin-$1/2$ chains with periodic boundary conditions and an odd number of sites, where the incompatibility between local N\'eel order and global topology forces the presence of frustrated loops whose length diverges in the thermodynamic limit~\cite{Dong2016}. 
As a result, the system is compelled to host a pair of parallel spins forming a domain wall that separates two locally ordered regions. 
The corresponding kink state represents the minimal-energy configuration compatible with the topology and constitutes the fundamental excitation characterizing the topologically frustrated phase.

Topological frustration is a remarkably robust phenomenon. 
Its effects persist across different dimensionalities~\cite{catalano2025newrungladderexploring} and can coexist with other sources of frustration, such as competing interactions in models like the ANNNI chain~\cite{Torre_2025}. 
Recent studies have further shown that TF systems display distinctive physical properties that make them particularly appealing from the perspective of quantum technologies~\cite{PRXQuantum.5.030319, catalano2025experimentalpreparationwstates}. 
In particular, the presence of a single delocalized topological excitations leads to ground states with a highly nontrivial structure, closely resembling multipartite entangled states such as W states~\cite{Maric2022_fate}.

These observations naturally motivate the characterization of topologically frustrated systems through the lens of \emph{quantum resources}. 
Among the various quantifiers explored so far~\cite{catalano2025newrungladderexploring, Giampaolo2019, Torre2023, Odavic2023, dl7m-8qyt, 10.21468/SciPostPhysCore.8.4.078}, entanglement and magic play a central role. 
Entanglement has long been recognized as a fundamental ingredient in quantum many-body physics and quantum information theory~\cite{Amico2008, Eisert2010}. 
Quantifiers such as the entanglement entropy have been extensively employed to detect and characterize quantum phase transitions, often exhibiting universal scaling behavior and logarithmic corrections at criticality~\cite{Sachdev2011}. 
Nevertheless, entanglement alone does not fully capture the features that distinguish quantum systems from their classical counterparts~\cite{Susskind2014, Chamon2014}. 
In particular, families of highly entangled states, such as stabilizer states generated by Clifford circuits~\cite{Aaronson2004}, can still be efficiently simulated on classical computers, as established by the Gottesman--Knill theorem~\cite{Gottesman1998}.

A genuine quantum advantage therefore requires resources that go beyond entanglement. 
This additional layer of quantumness is captured by the notion of \emph{magic}, which quantifies the non-stabilizer character of quantum states and operations~\cite{Bravyi2005, Liu2022, Veitch2014}. 
Originally introduced in the context of fault-tolerant quantum computation~\cite{Howard2017}, magic has recently emerged as a powerful diagnostic tool in many-body physics, providing insights into quantum chaos, thermalization, and critical phenomena~\cite{Haug2023, Liu2024, Bharti2022, Regula2022}. 
More recently, magic has also entered the domain of high-energy physics, where magic quantifiers such as the mana have been linked to scrambling, operator growth, and the emergence of chaos in holographic and black-hole-related settings~\cite{Goto2022, Garcia2023, White2021}.

The aim of this work is to present a unified perspective on topologically frustrated quantum systems by analyzing their properties in terms of quantum resources. 
In particular, we show how the peculiar ground-state structure induced by topological frustration leads to additive and clearly identifiable contributions to both entanglement and magic, which can be directly traced back to the presence of delocalized topological excitations. 
To this end, in Sec.~\ref{sec:GSStruncture} we analyze the ground-state structure of TF systems in the vicinity of the classical point, highlighting the emergence of W-state-like components. 
In Sec.~\ref{sec:EE} we investigate the behavior of entanglement-related quantifiers, demonstrating the decoupling between frustration-induced and non-frustrated contributions. 
In Sec.~\ref{sec:magic} we extend this analysis to magic, quantified through non-stabilizer R\'enyi entropies, and show that analogous additive structures persist throughout the phase diagram. 
Finally, conclusions and future perspectives are discussed in Sec.~\ref{sec:conclusions}.

\begin{figure}[t!]
	\centering
	\includegraphics[width=.8\columnwidth]{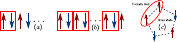}
	\caption{The mechanism underlying topological frustration can be understood through the following steps: (a) impose AFM couplings, which naturally favour a two-site periodic N\'eel configuration; (b) disrupt this periodic pattern by selecting an odd number of spins, thereby preventing the N\'eel order; (c) finally, close the system into a ring by enforcing PBCs, which forces the incompatibility introduced in (b) to extend throughout the entire system and gives rise to topological frustration.}
	\label{fig:Top_fr}
\end{figure}

\section{Ground state structure with topological frustration in quantum spin systems}
\label{sec:GSStruncture}

In this section, we illustrate the basic mechanism underlying the presence of TF within a quantum system and the resulting phenomenology. 
As anticipated in the Introduction, this mechanism is rather general. 
For illustrative purposes, we therefore focus on a paradigmatic model, namely the one-dimensional XYZ spin chain of length $N$, subjected to a uniform magnetic field along the $z$ axis~\cite{catalano2025experimentalpreparationwstates,Franchini2017}:
\begin{equation}
	H_{\mathrm{XYZ}} = \sum_{n=1}^{N} \sum_{\alpha=x,y,z}
	J_{\alpha}\, \sigma_{n}^{\alpha} \sigma_{n+1}^{\alpha}
	+ h \sum_{n=1}^{N} \sigma_{n}^{z}. \label{XYZ}
\end{equation}
Here $\sigma^{\alpha}$ ($\alpha = x, y, z$) denote the Pauli matrices, $h$ is the external magnetic field, and $J_{\alpha}$ represent the nearest-neighbor interaction strengths along the corresponding spin directions. 
Topological frustration is induced by imposing frustrated boundary conditions (FBCs), namely periodic boundary conditions ($\sigma^\alpha_j = \sigma^\alpha_{j+N}$), antiferromagnetic interactions, and an odd number of spins, $N = 2M + 1$ with $M \in \mathbb{N}$ (see Fig.~\ref{fig:Top_fr}).

To isolate the essential features of TF and eliminate model-specific quantum fluctuations, we begin by analysing the system at its classical point, defined by the parameter choice $(J_x = 1,\; J_y = J_z = h = 0)$, for which all terms in the Hamiltonian commute. 
At this point, simple geometric arguments~\cite{Torre2023} show that the ground-state manifold is $2N$-fold degenerate. 
These ground states correspond to the kink configurations illustrated in Fig.~\ref{fig:Top_fr}, obtained by placing a single pair of parallel spins, either both up or both down, at each of the possible positions along the chain.
These $2N$ kink states can be organized into two distinct families characterized by different magnetization parities. 
Within each family, any state can be obtained from another by a spatial translation, corresponding to a shift of the kink position along the chain. 
An explicit representation of these states is given by
\begin{equation}\label{kinks}
	\ket{k}  = \mathcal{T}^{k-1} \bigotimes_{j=1}^{M} \sigma_{2j}^{z} \ket{-}^{\otimes N}, \qquad
	\ket{k'} = \mathcal{T}^{k-1} \bigotimes_{j=1}^{M} \sigma_{2j}^{z} \ket{+}^{\otimes N}.
\end{equation}
where $\mathcal{T}$ denotes the lattice translation operator.

A distinctive consequence of topological frustration is thus the emergence of a ground-state degeneracy that grows linearly with the system size. 
When quantum fluctuations are reintroduced by moving away from the classical point, this degeneracy is lifted. 
Perturbation theory in the vicinity of the classical regime~\cite{Maric2022_fate,Sen2008,Catalano2022} shows that the $2N$-fold degenerate manifold reorganizes into a low-energy band with a gapless dispersion. 
This behavior sharply contrasts with the gapped phase characterizing the same region of parameter space in the corresponding unfrustrated systems.

The eigenstates forming the lowest-energy band can be written as coherent superpositions of the classical kink configurations in Eq.~\eqref{kinks},
\begin{align}
	\label{eq:wkinkstate}
	\ket{\omega_{p}} = \frac{1}{\sqrt{2N}} \sum_{k=1}^{N} e^{i p k} \left( \ket{k} + \ket{k'} \right),
\end{align}
where $p$ denotes the quantized momentum, $p = 2\pi \ell / N$, with $\ell = -\tfrac{N-1}{2}, \ldots, \tfrac{N-1}{2}$.

The defining feature of the states in Eq.~\eqref{eq:wkinkstate} is the presence of a number of coherently superposed components that scales linearly with the system size. 
This delocalized structure has profound consequences for the quantum properties of topologically frustrated systems. 
In the following section, we show how it gives rise to characteristic contributions to entanglement-related quantifiers, thereby providing a clear link between topological frustration, delocalized excitations, and quantum correlations.

\begin{figure}[t!]
	\centering
	\includegraphics[width=.8\columnwidth]{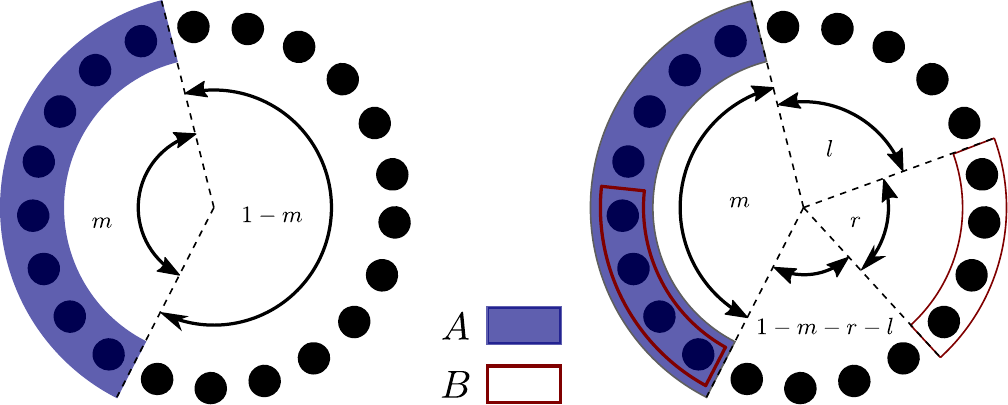}
	\caption{Schematic representation of two different entanglement partitions.
		(Left) Bipartition of the system for the entanglement entropy. The system is divided into two complementary subsystems $A$ made of $M=m N$ spins and $B$ made of $N-M=N(1-m)$ spins. 
		(Right) Disconnected entanglement entropy, defined between two disjoint subsystems $A$ and $B$ embedded in a larger system, which captures non-local quantum correlations beyond the area law contribution. The lengths $n$, $m$, and $l$ are normalized with respect to the length of the system.}
	\label{fig:EE_partitions}
\end{figure}

\section{Entanglement properties with TF}
\label{sec:EE}

As a measure of entanglement we start by considering the \emph{entanglement entropy} (EE). (Bi-partite) EE is a central quantity in quantum many-body physics and quantum information theory, providing a measure of quantum correlations between two complementary parts of a system. We begin by recalling its definition. Given a pure state $\ket{\psi}$ and a bipartition of the system ($\mathcal{H}=\mathcal{H}_A\otimes\mathcal{H}_B$), the entanglement entropy with respect to the subsystem $A$ is defined as the von Neumann entropy of its reduced density matrix $\rho_A=\mathrm{Tr}_B\ket{\psi}\!\bra{\psi}$~\cite{Amico2008,Eisert2010,Hastings2007,CalabreseCardy2004}. 
\begin{equation}\label{EE}
	S(A) \equiv -\mathrm{Tr}(\rho_A\log_2\rho_A).
\end{equation}
In one-dimensional spin chains EE is typically evaluated for the GS $\ket{\psi}$ by partitioning the system into two connected subsystems, $A$ and $B$, of lengths $M=m N$ and $N-M=N(1-m)$, respectively (see Fig.~\ref{fig:EE_partitions}).
It is then computed from the eigenvalues $\{\lambda_i\}$ of $\rho_A$ according to $S(A)=-\sum_i \lambda_i \log_2 \lambda_i$.
For ground states of local gapped Hamiltonians, the EE typically obeys an \emph{area law}, namely the bipartite von Neumann entropy $S(A)$ scales proportionally to the size of the partition's boundary $\partial A$, i.e., $S(A)=\mathcal{O}(|\partial A|)$, rather than its volume, up to sub-leading corrections~\cite{Eisert2010, Hastings2007, CalabreseCardy2004, Srednicki1993}. In particular, for one-dimensional spin chains, the area law implies that the EE of the ground state saturates to a constant value independent of the subsystem size, reflecting the short-range nature of quantum correlations in gapped phases~\cite{Amico2008,Eisert2010,Hastings2007}. Deviations from this behavior indicate the presence of long-range quantum correlations or critical phenomena~\cite{Amico2008,Eisert2010,Hastings2007,CalabreseCardy2004}.

Considering the $p=0$ case of Eq.~\eqref{eq:wkinkstate}, that is the zero-momentum GS of a TF chain near the classical point ($J_y=J_z=0, h \ll 1$), the expression of the EE in thermodynamic limit, was analytically computed in Ref.s~\cite{Giampaolo2019,Torre2023}, and it reads
\begin{equation}\label{EE_TF_classical}
	S^{\mathrm{TF}}(A) = 1 - m \log_2 m - (1-m) \log_2 (1-m).,
\end{equation}

This relation constitutes our first significant finding. Indeed, the first term in Eq.~\eqref{EE_TF_classical} corresponds to the area-law contribution, which is the same term appearing in the unfrustrated version of the model and reflects the $\mathbb{Z}_2$ symmetry of the model.
The size-dependent terms arise from the presence of TF and constitute a peculiar violation of the expected area law. This effect, although directly linked to the extensive degeneracy of the ground state manifold at the classical point, does not lead to a divergence of the entanglement entropy for large subsystems, since the entropy contribution reaches a finite maximum value for $m \rightarrow 1/2$. Note, however, that in an infinite chain the EE is always increasing for growing partitions lengths. 
In fact, this entropic contribution to the EE is that of a single delocalized excitation, which does not possess any intrinsic length scale, except for the total system size~\cite{Giampaolo2019}. This single particle contribution to the EE remains evident in the whole frustrated phase ($h<1$ for the $XYZ$ Hamiltonian in Eq.~\eqref{XYZ}) where the EE can be written as
\begin{equation}\label{EE_TF_phase}
	S^{\mathrm{TF}}(A) = S^{\mathrm{NF}} (A) - m \log_2 m - (1-m) \log_2 (1-m),
\end{equation}
where $S^{\mathrm{NF}} (A)$  is the area law limit obtained in absence of TF~\cite{Giampaolo2019}. This structure seems to be shared by all entropic quantum resources: as argued in \cite{dl7m-8qyt}, any quantum resource measured by an entropy ${\cal O} (\Omega)$, depending on a given set of parameters $\Omega$, for a TF system in the thermodynamic limit can be written as
\begin{equation}
   {\cal O}^{\mathrm{TF}} (\Omega) = {\cal O}^{\mathrm{NF}} (\Omega) + {\cal O}^{\mathrm{e}} (\Omega) \: ,
   \label{eq:ResDecomp}
\end{equation}
where ${\cal O}^{\mathrm{NF}} (\Omega)$ is that resource for the corresponding non frustrated system, while ${\cal O}^{\mathrm{e}} (\Omega)$ is the contribution due to the excitation in the ground state, which can be calculated (often analytically) in the vicinity of the classical point using the kink-state superposition in Eq.~\eqref{eq:wkinkstate} and remains constant throughout the phase.
Although no analytical proof of this relation is currently available, there is strong numerical evidence supporting this behaviour~\cite{Torre_2025,catalano2025newrungladderexploring,dl7m-8qyt,10.21468/SciPostPhysCore.8.4.078}. In the following, we examine the validity of Eq.~\eqref{eq:ResDecomp} in the context also of non-stabilizerness.

To better appreciate the contribution of the delocalized excitation to the physics of the system using the lens of entanglement, a more informative quantity is the \emph{disconnected entropy}~\cite{Zeng2015,Fromholz2020}
\begin{equation}\label{DEE}
	S_\alpha^D = S_{\alpha} (A) + S_{\alpha} (B) - S_{\alpha} (A\cup B) - S_{\alpha} (A\cap B),
\end{equation}
where the $B$ partition has to be chosen not contiguous, and with one of its two subsystem included in the partition $A$ (see Fig.~\eqref{fig:EE_partitions}). Here, we choose to employ the $\alpha$–R\'enyi entropies $S_\alpha (A) \equiv \frac{1}{1-\alpha} \log_2 \big[ \mathrm{Tr}(\rho_A)^\alpha \big]$, which generalize the von Neumann entropy and tends to it in the $\alpha \to 1$ limit.
The combination in Eq.~\eqref{DEE} is chosen so as to eliminate the possible contributions from both the area and the volume-law (which scale linearly with the subsystem size), and extract the genuine log-range entanglement (LRE)~\cite{Fromholz2020,Micallo2020}. In Ref.~\cite{Torre2023} the authors compute Eq.~\eqref{DEE} for the state in Eq.~\eqref{eq:wkinkstate} with $p=0$, finding in thermodynamic limit the expression
\begin{align}\label{DEE_near_zero}
	\tilde{S}_2^D (m,l) & = -\left[\log_2\left(m^2+(1-m)^2\right)-\log_2\left((1-\dfrac{m}{2})^2+\left(\dfrac{m}{2}\right)^2\right) \right. \nonumber \\  & \left.+\log_2\left(l^2+(1-l-m)^2+\dfrac{m^2}{2}\right)-\log_2\left(l^2+\left(1-l-\dfrac{3m}{2}\right)^2+\dfrac{5}{4}m^2\right)\right],
\end{align}
where $m$ and $l$ are the the normalized length of the partitions as in Fig.~\ref{fig:EE_partitions}. TF therefore induces genuinely long-range entanglement throughout the chain, generated by a delocalized, fractional, topological excitation, whose correlation length spans the entire system.

Most importantly, in accord with the arguments of \cite{dl7m-8qyt} exposed above, these considerations extend over the whole phase diagram~\cite{Torre2023}. Thus, Eq.~\eqref{DEE_near_zero} remains valid for any value of the Hamiltonian parameters in Eq.~\eqref{XYZ}, as long as the dominant interaction is AFM and $h<1$. This means that DEE signals the presence of a stable topological excitation induced by TF, beyond the simple kink-state superposition and in a non-integrable model.

\begin{figure}[t!]
	\centering
	\includegraphics[width=.9\columnwidth]{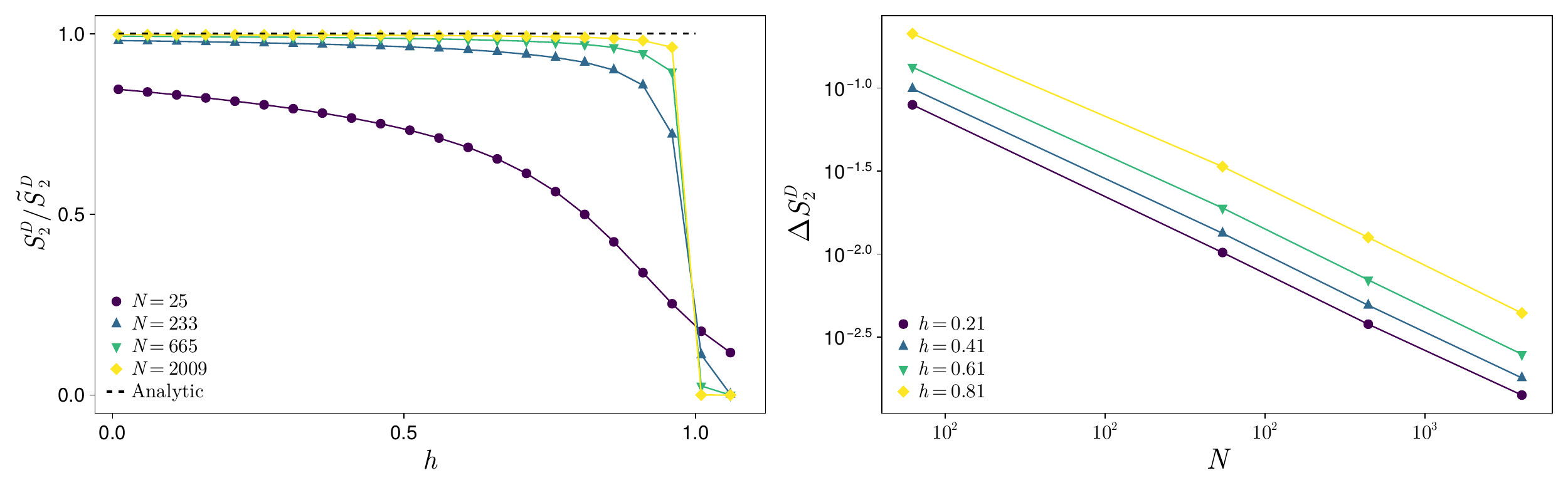}
	\caption{Disconnected $2$-R\'enyi entanglement entropy in the TF phase. (Left) Normalized $S^{D}_{2}(m,l)$ with respect to the value in Eq.~\eqref{DEE_near_zero}, plotted as a function of the magnetic field $h$ for different system sizes. The partition lengths are fixed to $m = (N-1)/2$, $l = (N-1)/8$, and $r =(N-1)/4$ (see Fig.~\ref{fig:EE_partitions}).  (Right) Scaling of 
    $\Delta S^D_2 (N) = \tilde{S}^D_2 (1/2,1/8) - S^D_2 (N, 1/2, 1/8, 1/4)$ 
    as a function of the chain length $N$. In the thermodynamic limit, $\Delta S_2^D$ vanishes as a power law, $\Delta S^{(2)}_{\mathrm{D}} = a N^{b}$, with exponent $b \simeq -0.935 \pm 0.006$, independently of $h$. 
	}
	\label{DDE_Isinf_scaling}
\end{figure}

To show this, we consider for instance the DEE in Eq.~\eqref{DEE} for the case $J_{y}=J_{z}=0, h \neq 0$, which corresponds to the frustrated transverse-field Ising chain. In this regime, the system becomes frustrated when the Ising interaction $\sigma^x_j \sigma^x_{j+1}$ dominates over the magnetic field term, namely for $\lvert h \rvert < 1$.  Although the calculation cannot be carried out analytically, an efficient numerical procedure is available~\cite{Torre2023}. The corresponding results are reported in Fig.~\ref{DDE_Isinf_scaling} for the entire TF phase $0 < h < 1$ and for specific choices of the subsystem lengths. The DEE is non-zero throughout the entire phase, clearly signalling the presence of TF. Moreover, the TF contribution to the entanglement attains the same value in the thermodynamic limit across the whole phase. The residual differences observed at finite system size stem from the finite correlation length~\cite{Torre2023}. Indeed, as the system approaches the phase transition, the increase in the correlation length slows down the convergence toward the asymptotic value. This convergence can be appreciated even more clearly by examining the difference between the 
analytical prediction in Eq.~\eqref{DEE} and the numerically obtained finite-size value as a 
function of the chain length $N$ (see the right panel of Fig.~\ref{DDE_Isinf_scaling}).  
TF is therefore not only present throughout the entire phase, but its contribution is also the same 
independently of the specific microscopic features of the model.  
It is important to note that this robustness against microscopic details strongly suggests a form 
of topological origin~\cite{Torre2023}.

\section{Non-stabilizerness of TF}
\label{sec:magic}

The analysis of the entanglement entropy already indicates that topological frustration substantially enhances the quantum nature of the ground state.
In this section, we explore this aspect 
through the lens of \emph{non-stabilizerness} (also known as \emph{magic}), a fundamental quantum resource required to achieve genuine quantum advantage~\cite{Preskill2018quantumcomputingin,NielsenChuang2000,10.5555/2011686.2011688,PhysRevLett.128.050402}.

The notion of magic arises from the need to identify which quantum resources enable genuinely quantum advantages beyond what can be efficiently simulated by classical means. In particular, it is well known that a broad class of quantum states, the so-called \emph{stabilizer states}, although they can exhibit substantial entanglement, can be efficiently simulated on a classical computer using only Clifford circuits, as formalized by the Gottesman--Knill theorem~\cite{Gottesman1998}. Magic quantifies the extent to which a quantum state or operation departs from this efficiently simulable set. In this sense, it captures a form of quantumness that is complementary to entanglement. 

Several approaches have been developed for its quantification. Prominent examples include the minimum and maximum relative entropies of magic, robustness of magic, Bell magic, stabilizer nullity, and the \textit{Stabilizer R\'enyi Entropies} (SREs)~\cite{Liu2022,PhysRevLett.128.050402, Haug2023_3,Bravyi2019,Pashayan2015}.  
These measures are generally computationally hard, requiring, for instance, optimization over all stabilizer decompositions or the evaluation of exponentially many correlators. Among all these quantities, SREs have attracted particular attention, especially due to the existence of efficient algorithms for their computation in specific settings such as tensor network states~\cite{Tarabunga2024criticalbehaviorsof,PhysRevLett.131.180401,10.21468/SciPostPhys.19.4.085,PRXQuantum.4.040317,kozic2025computingquantumresourcesusing,jpt3-xjjf}. Because tensor network representations work well in spin chains, even for TF ones, SRE is the measure that we adopt throughout this work.

To introduce SREs, we start by defining the set of all possible Pauli strings $\mathcal{P}_N=\bigotimes_{i=1}^N P_i$, with $P_{i} \in \{ \mathbb{I}, \sigma_{i}^{x}, \sigma_{i}^{y}, \sigma_{i}^{z} \}$, obtained as subset of $\mathcal{G}_N$ without considering the possible phase factors. The stabilizer states have the property that are stabilized by exactly $2^N$ Pauli strings~\cite{Aaronson2004}. We can then resort to the Pauli spectrum, defined as $\textrm{spec}(\ket{\psi}) = \{ \langle \psi | P | \psi \rangle \,|\, P \in \mathcal{P}_N \}$, that is, the set of expectation values of all Pauli strings, to quantify the degree of non-stabilizerness of a given quantum state. In fact, for a stabilizer state there will be exactly $2^N$ Pauli strings satisfying $\ev{P}{\psi}^{2q} = 1$, with $q$ even, with all other elements of the Pauli spectrum being zero. Consequently, the ratio $(1/2^N) \sum_P \ev{P}{\psi}^{2q} = 1$ if and only if the state is a stabilizer state. It is then possible to define the \emph{Stabiliser R\'{e}nyi entropies} family of functions
\begin{equation}\label{SRE}
	\mathcal{M}_q = \dfrac{1}{1-q}\log_2(\zeta_q),\qquad \zeta_q\equiv \frac{1}{2^N} \sum_{P\in \mathcal{P}_N}\ev{P}{\psi}^{2q},
\end{equation}
for $q\geq 2$. Intuitively, the measures defined in Eq.~\eqref{SRE} quantify the spread of the Pauli spectrum of a given state, interpreted as a probability distribution. The case $q=2$ is the most commonly used; therefore, in what follows, we focus on this instance and use the acronym SRE to refer specifically to the $q=2$ case.

Equipped with the measure in Eq.~\eqref{SRE}, we now compute the magic of GHZ and W states, as this will be important for the subsequent discussion. GHZ and W states constitute two paradigmatic families of multipartite entangled states with key relevance to quantum technologies.  
GHZ states enable Heisenberg-limited quantum metrology and phase estimation, achieving $1/N$ scaling beyond the shot-noise limit~\cite{Giovannetti2006,Giovannetti2011}, and serve as essential resources for quantum nonlocality tests and multi-party cryptographic protocols such as quantum secret sharing~\cite{Hillery1999}.  W states, by contrast, are notable for their robustness to qubit loss and decoherence, retaining bipartite entanglement even after subsystem removal~\cite{Dur2000}. This resilience makes them valuable for quantum networking and distributed quantum sensing, with various experimental demonstrations, such as in ion traps~\cite{Haefner2005}. But the structure of W states provides valuable resources for several tasks, from teleportation of unknown states~\cite{Joo2003}, to quantum secret sharing and key distribution~\cite{Wang2007}, from random number generation~\cite{grafe2014}, to quantum thermodynamics~\cite{daug2016}, from photonic and atomic encoding schemes.~\cite{Ebert2015,Vijayan2020}, to initial resource in Grover-like quantum searches~\cite{Grover1997} restricted to specific subspaces~\cite{Biham2002}

It is straightforward from the definition of GHZ states for N qubits
\begin{equation}
	\ket{\text{GHZ}_N} = \frac{1}{\sqrt{2}} \left( \ket{\uparrow}^{\otimes N} + \ket{\downarrow}^{\otimes N} \right),
\end{equation}
where $\ket{\uparrow}$ and $\ket{\downarrow}$ are the eigenvector of the Pauli $\sigma^z$ matrix, and from the measure definition Eq.~\eqref{SRE}, to verify that GHZ states possess zero non-stabilizerness. As an illustrative example, let us consider the two-qubit case 
$\ket{\text{GHZ}_2} = (\ket{\downarrow \downarrow} + \ket{\uparrow \uparrow}) / \sqrt{2}$.  A direct calculation shows that this state is an eigenstate of the following $2^2 = 4$ Pauli operators: 
$\mathbb{I} \otimes \mathbb{I}$, $\sigma^x \otimes \sigma^x$, $-\sigma^y \otimes \sigma^y$, and $\sigma^z \otimes \sigma^z$, 
all with eigenvalue $+1$, and from Eq.~\eqref{SRE}, $\mathcal{M}_2=0$. GHZ states, therefore, although useful for different quantum tasks, cannot constitute a reliable resource for achieving quantum advantage.

On the contrary W states posses a certain amount of magic. Indeed let us consider the even more general class of translational invariant \emph{generalized W states}~\cite{10.21468/SciPostPhysCore.8.4.078}
\begin{equation}
	\label{eq:Wpstate}
	\ket{W_{p}}=\frac{1}{\sqrt{N}}\sum_{j=1}^N e^{\imath p j} \sigma^z_j\ket{-}^{\otimes N}.
\end{equation}
In this expression, the states $\ket{\pm}$ represent the eigenstates of the Pauli operator $\sigma^x$ associated with the eigenvalues $\pm 1$ respectively, while $N$ indicates the total number of qubits in the system and $p$ is a phase shift that can be interpreted as the momentum of the state. 
It is possible to prove~\cite{10.21468/SciPostPhysCore.8.4.078} that the expression of the SRE for the generalized $W$-states is given by
\begin{align}
	\mathcal{M}_{2} (p,N) = - \log_2{\left( - \dfrac{11 - 12N + \frac{\sin{\left( (2 - 4N) p \right)}}{\sin{(2 p )}}}{2 N^{3}}\right)}. \label{TDexpressionFinite}
\end{align}
where the momenta $p$ obeys to the quantization rules of the states in Eq.~\eqref{eq:wkinkstate}.
The minimum is obtained in the limit $p \to 0$, corresponding to the standard W states:
\begin{align}
	\mathcal{M}_{2} (0,N) = 3 \log_{2}{(N)} - \log_{2} {(7 N - 6)}. \label{TDexpression}
\end{align}
If we impose periodic boundary conditions to the state in Eq.~\eqref{eq:Wpstate} and thus quantize $p$ as $p = \tfrac{2\pi \ell}{N}$, where $\ell$ is an integer ranging from $-\tfrac{N-1}{2}$ to $\tfrac{N-1}{2}$, the momentum dependence disappears from the SRE in Eq.~\eqref{TDexpressionFinite} and for $\ell \ne 0$ we have
\begin{align}
	\left.\mathcal{M}_{2} \left( \frac{2 \pi}{N} \ell,N \right)\right|_{\ell\neq0}  \!\!\!\!\!\!\!\!=\mathcal{M}_{2} (0,N) + \log_2{\left(  \dfrac{7 N - 6}{6 N - 6} \right)}.
	\label{eq:extramagic}   
\end{align}
Thus, we see that resourceful states in Eq.~\eqref{eq:Wpstate} belong to a different class of magic compared to GHZ states. 

\begin{figure}[t!]
	\centering
	\includegraphics[width=.7\columnwidth]{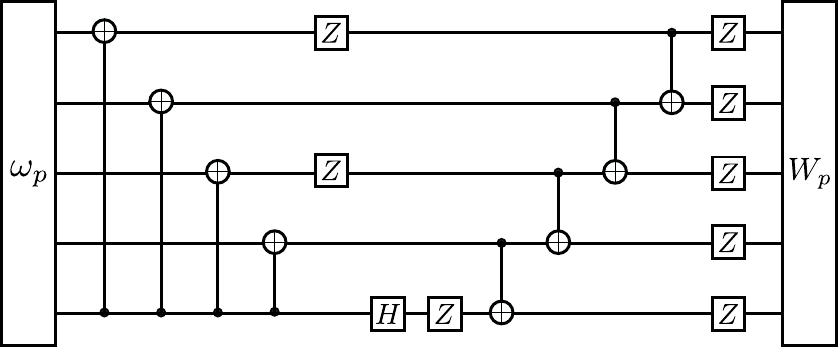}
	\caption{Pictorial representation of the Clifford circuit $\hat{\mathcal{S}}$ in Eq.~\eqref{cliffordcircuit} for $5$ qubits.  
		The boxes labeled $H$ and $Z$ denote the Hadamard gate and the $\sigma^{z}$ operator acting on the corresponding qubit.  
		CNOT gates are depicted as lines connecting a filled black dot, indicating the control qubit, to a circle marking the target qubit.
	}
	\label{fig:Clifford_circuit}
\end{figure}

We now turn to the computation of the non-stabilizerness for the superposition of kink states in Eq.~\eqref{eq:wkinkstate}.
It is sufficient to note that these states can be obtained from those in Eq.~\eqref{eq:Wpstate} by applying the Clifford circuit~\cite{Odavic2023},  
\begin{equation}
	\hat{\mathcal{S}} = 
	\prod_{j=1}^{N-1} \mathbf{C}(N,N-j)
	\left(\prod_{j=1}^{\lfloor N/2 \rfloor} \sigma_{2j-1}^z \right)
	\mathbf{H}(N)\sigma_N^z
	\prod_{j=1}^{N-1} \mathbf{C}(j,j+1)
	\Pi^z ,
	\label{cliffordcircuit}
\end{equation}
where $\mathbf{H}(j) \equiv \frac{1}{\sqrt{2}}(\sigma^x_j + \sigma^z_j)$ denotes the Hadamard gate acting on the $j$-th qubit, 
$\mathbf{C}(j,l) \equiv \exp\!\left[i\frac{\pi}{4}(1-\sigma^x_j)(1-\sigma^z_l)\right]$ represents the CNOT gate on the $l$-th qubit controlled by the $j$-th one, 
and $\Pi^z = \bigotimes_{j=1}^{N}\sigma^z_j$ is the parity operator along the $z$ direction (see Fig.~\ref{fig:Clifford_circuit}). Since the states $\lvert \omega_p \rangle$ are obtained from the $\lvert W_p \rangle$ through a Clifford circuit, they possess the same amount of magic. The ground states of TF models thus fall into the same non-stabilizerness class as (generalized) $W$ states, 
making them resourceful states for quantum technologies. 
Motivated by this observation, a recent experimental protocol for their preparation has been 
proposed and demonstrated on the QuEra digital quantum computing platform~\cite{catalano2025experimentalpreparationwstates}.

\begin{figure}[b!]
	\centering
	\includegraphics[width=.9\columnwidth]{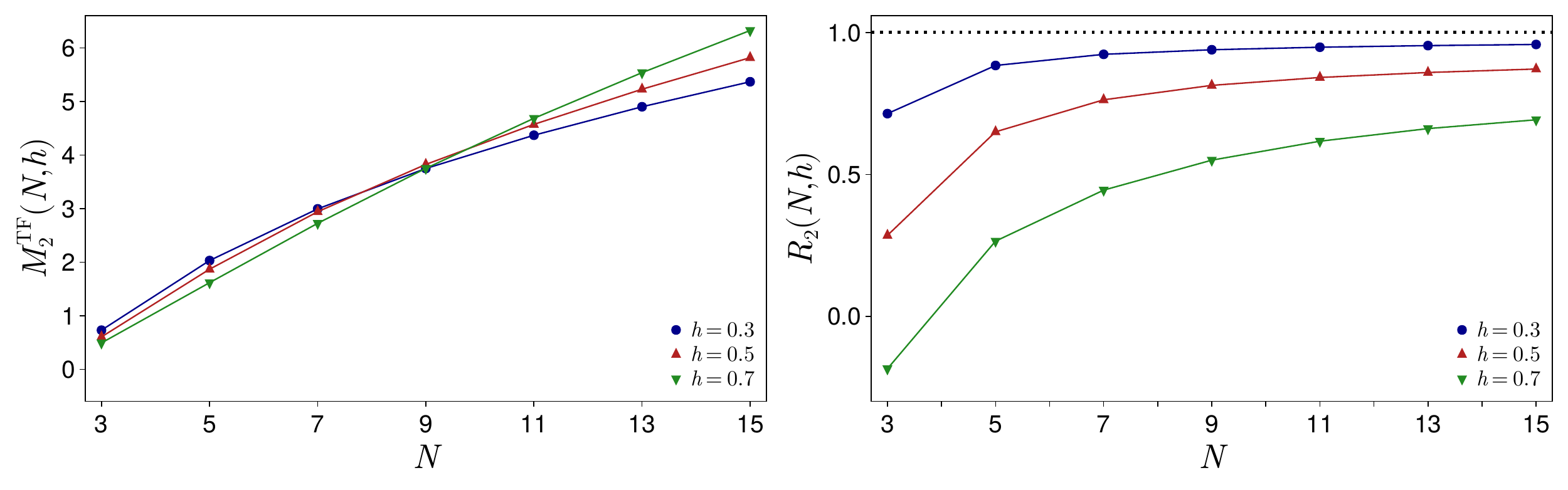}
	\caption{Stabilizer $2$-R\'enyi entropy for the frustrated Ising chain ($J_y = J_z = 0$) in the frustrated regime $h < 1$. (Left) At variance with the SRE for the unfrustrated model, that displays a strictly linear growth with the system size, TF induces an additional non-linear contributions. (Right) The relative frustrated SRE correction in Eq.~\eqref{eq:ratio} for $p=0$ as a function of the chain length $N$ for different values of $h$. In the thermodynamic limit, it approaches unity, indicating that the SRE of the system can be written as the sum of two terms, one due to the presence of the excitation and the other one due to presence of the magnetic field.
	}
	\label{magic_Ising_scaking}
\end{figure}

Moving away from the classical point, the scaling of the $2$-SRE Eq.~\eqref{SRE} for the case of the frustrated Ising chain ($J_y=J_z=0$) is reported in Fig.~\ref{magic_Ising_scaking} for different values of the magnetic field. The data exhibit a distinctly non-linear dependence on $N$. Such non-linearity is a direct manifestation of topological frustration, as it is well established~\cite{PhysRevA.106.042426} that the SRE of the unfrustrated Ising model grows linearly with the system size. The numerics then suggests that, in the thermodynamic limit, the SRE can be decomposed into two contributions: a local term identical to that of the corresponding unfrustrated model, and an additional term arising from the delocalized excitation. Consequently, for sufficiently large $L$, one expects that in the frustrated phase it is possible to write $\mathcal{M}_2^{\textrm{TF}} = \mathcal{M}_2^{\textrm{NF}} + \mathcal{M}_2^e$, where $\mathcal{M}_2^e = \mathcal{M}_{2} (0,N)$ from Eq.~\eqref{TDexpression}.
We notice that, for this model, the unfrustrated counterpart is obtained by reversing the sign of the interaction ($J_x=1$) while keeping both the chain length $N$ and the value of the magnetic field $h$ fixed. 

To better show this decomposition, in the right panel of Fig.~\ref{magic_Ising_scaking} we show the behavior of the \textit{relative frustrated SRE correction}, defined as the difference between the SRE of the frustrated and unfrustrated models for fixed $h$ and $N$, normalized by the corresponding value of the SRE near the classical point from Eq.~\eqref{TDexpression}
\begin{equation}
	\label{eq:ratio}
	R_2 = 
	\frac{\mathcal{M}_2^{\textrm{TF}} - \mathcal{M}_2^{\textrm{NF}}}{\mathcal{M}_{2} (p,N)} \: ,
\end{equation}
with $p=0$. Across all cases displayed in the figure, the numerical results closely follow our theoretical expectations: in the frustrated regime ($h < 1$), the ratio in Eq.~\eqref{eq:ratio} approaches 1 as the system size grows.

\begin{figure}[t!]
	\centering
	\includegraphics[width=.9\columnwidth]{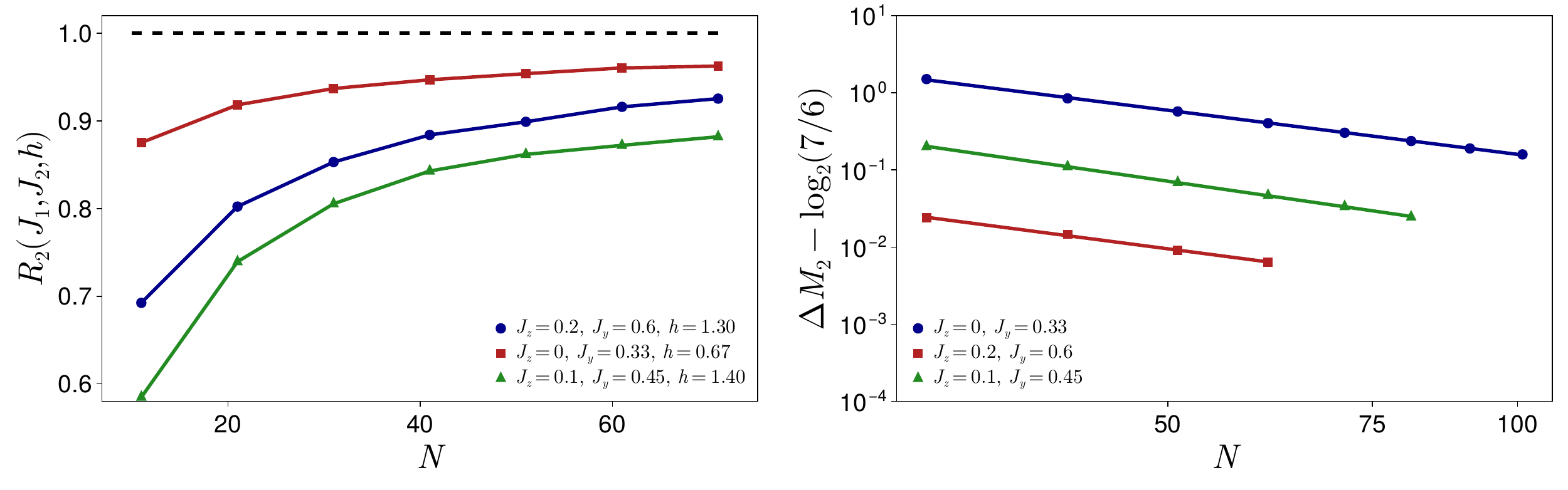}
	\caption{(Left) 
    The ratio $R_2$ defined, as a function of $N$ for different sets of parameters in the frustrated phase.  In all cases, we observe a power-law convergence to $R_2=1$. 
    (Rigth) Finite-size scaling analysis of the discontinuity in the SRE for different sets of anisotropies ($J_x=1$ in all cases ). It is possible to observe a power-law decay to the thermodynamic values that is equal to $\log_2(7/6)$.
 	}
	\label{XYZ_magic_plots}
\end{figure}

Let us now turning to the more general case of the XYZ chain ($J_y\neq 0, J_z \neq 0$). We will assume that $J_{x} = 1$, $\vert J_{y} \vert,\, \vert J_{z} \vert < J_{x} $, and, without loss of generality, that $J_{z} \ge -J_{y}$. 
While close to the Ising line the ground state is unique (and static), as shown in~\cite{Catalano2022,SaccoShaikh2024} with TF there exists a critical value of the external magnetic field $h^{*} > 0$ such that, for $\lvert h \rvert < h^{*}$, the ground-state manifold becomes two-fold degenerate and is spanned by states with finite, opposite momenta $p \neq 0$. Qualitatively, in this region the two dominant interactions are AFM and both get frustrated by the boundary condition: in order to reduce the energy cost of such frustrations, the ground state has to acquire some kinetic energy. Remarkably, even in this region the low-energy states retain the single excitation description, as generalizations (local dressing) of the kink-superposition in \eqref{eq:wkinkstate}, with a finite $p$. Thus, it is still possible to apply the decomposition in Eq.~\eqref{eq:ResDecomp} and evaluate the ratio in Eq.~\eqref{eq:ratio}, but using an appropriate $p$ in the denominator~\cite{10.21468/SciPostPhysCore.8.4.078}. The results, shown in the left panel of Fig.~\ref{XYZ_magic_plots}, indicate that also in this case the complexity of the ground state can be decomposed into the sum of a non-frustrated contribution and an additional term arising from the presence of the excitation.

The attentive reader might have noticed a finite jump between the SRE of a zero-momentum kink (see Eq.~\eqref{TDexpression}) and that of a finite momentum one in Eq.~\eqref{eq:extramagic}. Even though the change of ground state degeneracy at $h=h^{*}$ should configure such condition as a first order (boundary) phase transition, no discontinuities in the ground state energy (or any of its derivatives) can be measured, nor it was possible to identify a local order parameter that detects the transition in the thermodynamic limit. Interestingly, the ground state SRE serves this purposes and presents a finite jump crossing the transition at $h=h^{*}$ as shown in the right panel of Fig.~\ref{XYZ_magic_plots}. Since, as shown in \cite{10.21468/SciPostPhysCore.8.4.078}, the half-chain entanglement entropy is continuous across the phase transition, this singular behavior of SRE indicates its ability to detect elusive quantum phase transition that defy standard classification, especially those characterized by extensive order parameters.

\section{Conclusions and Outlook}
\label{sec:conclusions}

In this work, we examined the phenomenology of topologically frustrated spin chains through the combined perspective of their entanglement and non-stabilizerness features. To elucidate the underlying mechanisms, we began by analysing the physics of these systems in 
the vicinity of a distinguished point in the parameter space, the classical point. In this regime, we characterised the structure of the ground-state manifold and demonstrated, via its entanglement properties, that a topological excitation is superimposed on the unfrustrated ground state.  Moreover, we showed that the ground state of TF models lies in the same complexity class as generalized $W$ states, making it a potentially valuable resource for quantum-information tasks. Finally, we investigated the effects of TF across the entire frustrated phase, showing that, while the phenomenology remains robust, the TF-induced contribution to quantum resources factorizes from the model-dependent ones in the thermodynamic limit, see Eq.~\eqref{eq:ResDecomp}.

The phenomenology of TF systems is far richer than what has been presented here, Indeed, TF gives rise to a number of remarkable and nontrivial effects in the low-energy physics of these systems. A first remarkable phenomenon is the modification of the phase diagram and the behaviour of order parameters. TF models seem to defy the predictions of Landau’s theory: while being introduced by a suitable choice of boundary conditions, TF can deeply modify the phase diagram of quantum spin chains even in their gapped phases, where a finite correlation length should prevent boundary conditions from affecting large distance behaviors. For example, it has been shown that, as from one side TF can destroy order parameters~\cite{Maric2020_destroy}, from the other it is also able to induce phase transitions to phases which are characterized by incommensurate antiferromagnetic order~\cite{Maric2020_induced}. Moreover, in the context of exactly solvable $2$-Cluster Ising models, it was shown that TF can destroy order parameters (replacing them with new string orders) on both sides of a phase transition, consequently changing its nature~\cite{Maric2022_nature}.

TF models also present a strong dynamical response to local perturbations. In fact, the gapless band to which the GS belongs and that comprises an extensive number of states induced by TF has profound implications for the dynamical behaviour. Indeed, even weak perturbations can hybridize multiple low-energy states, thereby producing observable effects in dynamical quantities. This has been explicitly demonstrated through the study of the Loschmidt echo dynamics following a local quantum quench in the external magnetic field of a TF Ising chain~\cite{Torre2022}. By comparing the time-dependent behaviour of the Loschmidt echo for AFM rings with an even (unfrustrated) and an odd (frustrated) number of spins, it was shown that in the former case the Loschmidt echo fluctuations are very small, while in the latter, that is with TF, the amplitude of LE oscillations is larger by two orders of magnitude for large systems, as a consequences of the hybridization of states within the gapless band.  
Remarkably, these results suggest the counterintuitive possibility of distinguishing between a system containing an Avogadro number $N_A$ of sites from one with $N_A + 1$ particles.  
Although these findings were obtained in the specific context of the Ising chain, their validity is expected to extend to other TF models, since the underlying mechanism relies solely on the characteristic spectral structure shared by all topologically frustrated Hamiltonians.

Although we discussed the physics of TF models only considering the case of short range 1D spin chains, the phenomenology is more general.
For example, in the TF ANNNI chain, the extensive frustration arising from local competing interactions and the sub-extensive contribution of TF play fundamentally different roles in shaping the phenomenology of these models. In \cite{Torre_2025} it was shown that, once more, the entanglement entropy naturally separates into two distinct contributions: a part already present in the absence of TF and a TF-induced term associated with the presence of (not one but) two excitations in the ground state. These excitations behave independently except for the constraint that both must minimize their momentum while being unable to occupy the same momentum state. As a consequence, they develop correlations, typical of an effective fermionic nature. Another interesting case recently considered is the quasi two-dimensional setting of three-leg quantum Ising ladder with toroidal topology~\cite{catalano2025newrungladderexploring}. The results show that TF change the phase diagram, shifting the critical point relative to the unfrustrated case. Furthermore, the scaling of the entanglement entropy is consistent with the presence of three bosonic quasiparticles in the ground state. 

The study of TF systems is far from complete. A natural direction for future research is the extension of these investigations to fully two-dimensional TF models. On the technological side, TF chains have recently been proposed as promising platforms for quantum batteries, where they display markedly enhanced performance in terms of resilience, charging times, and energy storage capacity~\cite{PRXQuantum.5.030319}.  
Furthermore, TF chains exhibit a notable degree of robustness against 
noise~\cite{Torre2023,PhysRevB.103.014429,Odavic2023_random}, and their experimental realization on current NISQ devices~\cite{catalano2025experimentalpreparationwstates}, together with 
favourable scaling prospects as hardware continues to advance, positions them as compelling candidates for the development of future quantum technologies.

\ack{The authors would like to thank Vanja Marić (University of Ljubljana), Jovan Odavić (University of Naples “Federico II”), Sven Benjamin Kožić (Ruđer Bošković Institute, Zagreb), Alioscia Hamma (University of Naples “Federico II”), and Pierre Fromholz (University of Basel) for insightful discussions and for their valuable contributions to this work.}

\funding{AGC has received funding from the following organizations: European Union via ICSC - Italian Research Center on HPC, Big Data and Quantum Computing (NextGenerationEU Project No. CN00000013), project EuRyQa (Horizon 2020), Italian Ministry of University and Research (MUR) via: Quantum Frontiers (the Departments of Excellence 2023-2027); the World Class Research Infrastructure - Quantum Computing and Simulation Center (QCSC) of Padova University; Istituto Nazionale di Fisica Nucleare (INFN): iniziativa specifica IS-QUANTUM.
GT, SMG and FF acknowledges support from the project "Implementation of cutting-edge research and its application as part of the Scientific Center of Excellence for Quantum and Complex Systems, and Representations of Lie Algebras", Grant No. PK.1.1.10.0004, co-financed by the European Union through the European Regional Development Fund - Competitiveness and Cohesion Programme 2021-2027.
Moreover, FF acknowledges the support of the Croatian Science Foundation (HrZZ) through the project IP-2025-02-1667, Mining the Quantum: Frustration, Disorder, and Devices.}

\bibliographystyle{unsrt}
\bibliography{bibliography.bib}

\end{document}